\documentclass[12pt]{article} 
\usepackage{graphicx}

\begin{document}

\centerline{\textbf{ \large A PHYSICS SHOW PERFORMED } }

\medskip

\centerline{\textbf{ \large BY STUDENTS FOR KIDS:}}

\medskip

\centerline{\textbf{ \large From Mechanics to Elementary Particle Physics\footnote{
The following article has been submitted to The Physics 
Teacher. After it is published, it will be found at
http://scitation.aip.org/tpt/.}}}

\vspace{0.4cm}

\centerline{{Herbi K. Dreiner}}

\vspace{0.2cm}

\centerline{\textit{\small Physikalisches Institut, Universit\"at Bonn}}
\centerline{\textit{\small Nu{\ss}allee 12, 53115 Bonn, Germany}}

\centerline{\footnotesize February 1$^{\mathrm{st}},$ 2007}

\vspace{0.6cm}

\textbf{\small ABSTRACT:} 
{\small In this article, we describe an initiative at the University
of Bonn, where the students develop and perform a 2 hour physics show
for school classes and the general public. The show is entertaining
and educational and is aimed at children aged 10 and older. For the
physics students this is a unique experience to apply their knowledge
at an early stage and gives them the chance to develop skills in the
public presentation of science, in front of 520 people per show. We
have extended the activity to put on an elementary particle physics
show for teenagers. Furthermore, local high schools have picked up the
idea; their students put on similar shows for fellow students and
parents. We would be interested in hearing about related activities
elsewhere.}

\vspace{0.8cm}

Physics students spend the early part of their training attending
physics and mathematics lectures, solving problem sets and
experimenting in laboratory courses. The program is typically
intensive and fairly rigid. They have little opportunity to follow
their own curiosity or apply their knowledge. There have been several
attempts to address this deficiency. For example, Prof. Clint Sprott
at the University of Wisconsin, where the author was a graduate
student, hosts a physics show entitled ``The Wonders of Physics.$\!$''
(The show has continued since 1984, http://sprott.physics.wisc.edu/
$\!\!$wop.htm). Dressed up as a mixture of circus director and
magician and assisted by students, Prof.~Sprott presents entertaining
and educating experiments to a regularly packed audience of all age
groups \cite{sprott}. This was in turn inspired by the ``Chemistry is
Fun'' presentations of Prof.~Basam Shakhashiri, also from the
University of Wisconsin (http://scifun.chem.wisc.edu/).

\medskip 

In 2001, we launched a similar activity in Bonn. Our idea was to give
a group of undergraduates (about 25-30) from every incoming class
(about 150-170) the opportunity to put on a physics show for kids aged
10 and older. Thus the essential point of this initiative is that the
\textit{students} develop and perform the show themselves; it is
entirely in their control. The students have full access to our
collection of demonstration experiments, assisted and advised by
Michael Kortmann, who is in charge of the collection. They also
receive assistance from our workshops in the development and
construction of new experiments. They can thus follow their own
curiosity, choosing the experiments they find interesting. They spend
many hours understanding the experiments and considering how to best
explain them to children, without oversimplifying. The show is a
unique opportunity for the students to apply their knowledge at an
early stage by teaching kids in a dramatic environment and to develop
valuable skills in the public presentation of science. The show has
been a resounding success with the public, with a regularly packed
520-people auditorium. For us however, the Bonn physics show is first
and foremost an educational tool addressed at our physics students
\cite{german}.  The enthusiasm with which they have picked up the idea
has more than rewarded our efforts, and their talent at public
presentation is a delight to observe.  Furthermore, we now have a very
large pool of experienced and enthusiastic students with whom we have
developed further outreach activities. For example, we participate in
the so-called Bonn Science Night. This is a science fair organized
every other year in the main university builing, an 18th century
palace. This effort was initiated and organized entirely by the
students themselves.

\medskip

Every year in January, we recruit a new group of students in the
second year class. The students participate voluntarily and for no
credit. Together with students from previous years, Michael Kortmann
and I attend the initial meetings. We pass on our experience and give
the main idea of how the show is set up.  Starting in March, the
students largely organize the meetings themselves. The last 8-10 days
before the first show, the students rehearse daily. Michael Kortmann
gives technical assistance and students from previous years and I give
advice on the text and presentation. There are three performances in
September and a repeat with three more at the end of February. One
performance is always for school classes (coming from as far as
$100\,$km) and two on the weekend are for the general public.

\medskip

The show lasts two hours, including a 20 minute intermission, when the
children can try out and have a close look at experiments at the front
of the lecture hall, as well as in the extensive lobby. There are two
MC's (one woman and one man) to guide the audience through the
show. Two important ingredients were introduced by the students in the
first year.  \textit{(i)} Short ($<90$ sec) home-made entertaining
movies to introduce the various physics topics (\textit{e.g.}
mechanics, magnetism, low-temperatures) into which the show is
divided. Over the years these have become quite elaborate. They can be
viewed at http://www.physikshow.uni-bonn.de/
\textit{(ii)} Between explanations, the experiments are accompanied by
up-beat music, which creates a unique energetic atmosphere, very much
appreciated by the young audience and the local radio station, which
has twice broadcast live from the show.

\medskip

Two weeks ago we started putting the introductory films, as well as
films of experiments on YouTube. The most successful film shows a
light boat built out of aluminum foil floating in an aquarium filled
with sulphur hexafluoride, Fig. 1. Within the first 3 weeks it had
over 290,000 viewings, making it the fourth most viewed German science
and technology film ``all time''. The film can be viewed at
http://www.youtube.com/watch?v=1PJTq2xQiQ0. Some trailers can also be
viewed on YouTube under the search words ``Physikshow'' and ``Bonn''.
One film, for example, shows momentum conservation with a multiple
pendulum followed by four test car crashes all to Beethoven's
9$^{\mathrm{th}}$ symphony, Beethoven of course originating from Bonn.
Another film shows ball bearings bouncing in a simulation of kinetic
gas theory, followed by a sequence of photos relating to atomic and
particle physics, accompanied by upbeat music.

\medskip

Over the years the students have built quite a few experiments
themselves. It started with a simple apparatus to let pickles and
other vegetables glow, when an electrical current flows through them.
Later, these included a hoover craft with a chair and a vacuum cleaner
engine, a superconducting train, the ``Polar express'', which is shown
in Figs. 2 and 3, and a large tank to simulate a tsunami wave. The
students also built a large box ($\sim 2\, \mathrm{m}^3$) which
produces smoke rings which travel through the entire lecture hall.

\medskip

This is the basic structure of the show into which the students have
introduced a tremendous variety. The first two years, the show was
similar to those of Prof.~Sprott: a series of experiments were
presented and explained. In the third year, the students developed a
question and answer scheme, where one of the MC's asked questions and
the other explained the physics. This created a better flow and also
dramatically improved the explanations, since the students had to
imagine possible questions the kids might have. In the fourth year,
the students built a ``time-travel machine'' out of an old telephone
booth and a smoke machine, and the two MC's visited the Stone Age,
Aristotle, Newton, \textit{etc.} asking questions directly of the
greats. Last year the students staged the entire physics show as a
large computer game, where they had to solve various physics problems
to advance to higher levels. For the movies the two protagonists
wandered through a virtual computer world to advance to the next
level.

\medskip

The physics show has mainly involved classical physics occasionally
with some quantum effects. However, when the particle physics
laboratory CERN turned 50 in 2004, the first year group had
sufficiently advanced in their studies that we could put on an
elementary particle physics and cosmology show addressed at
high-school students. This was a huge success with the local schools.

\medskip

In March, 2006, we were honored that the Deutsche Museum, in
M\"unchen, invited us to put on three shows in the distinguished
Ehrensaal, something like the hall of fame of German science and
technology. This was an exciting opportunity for the students and our
first big away game. It was also a logistical challenge to get the
experiments to M\"unchen at a reasonable cost. The performances were
very well received and the Museum has suggested turning this into a
semi-annual event.

\medskip

The basic idea of the physics show also carries over to high-school
physics. Two local high-schools have independently put on shows for
their fellow students and their parents. At the Kollegium Josephinum,
the students put on a 60 min show on the topic of ``water''. At the
school in Hennef, they have a regular group called the ``Physikusse'',
which also puts on small shows. 
(http://www.ge-hennef.de/foefo/ $\mathrm{begabung}_{-}$physikusse.htm)

\medskip

In October, 2006, in honor of the tremendous amount of effort the
students have put into the physics show, they received the Alumni
Prize of the University of Bonn for student initiative.

\vspace{1cm}

\noindent\textbf{\Large Acknowledgments}

\medskip

\noindent It is a great pleasure to thank Michael Kortmann. We have 
been pursuing this project jointly from the beginning and it has
always been an honour to work with him. Furthermore, I would like to
thank our secretaries Patricia Z\"undorf, Dagmar Fassbender and Sandra
Heidbrink for their support and their efforts way beyond the call of
duty. Most of all I would like to thank the students of the past five
years for the great fun and enthusiasm they have brought to the Physikshow.


\centerline{\includegraphics[width=8.66cm,height=6.5cm,angle=0]{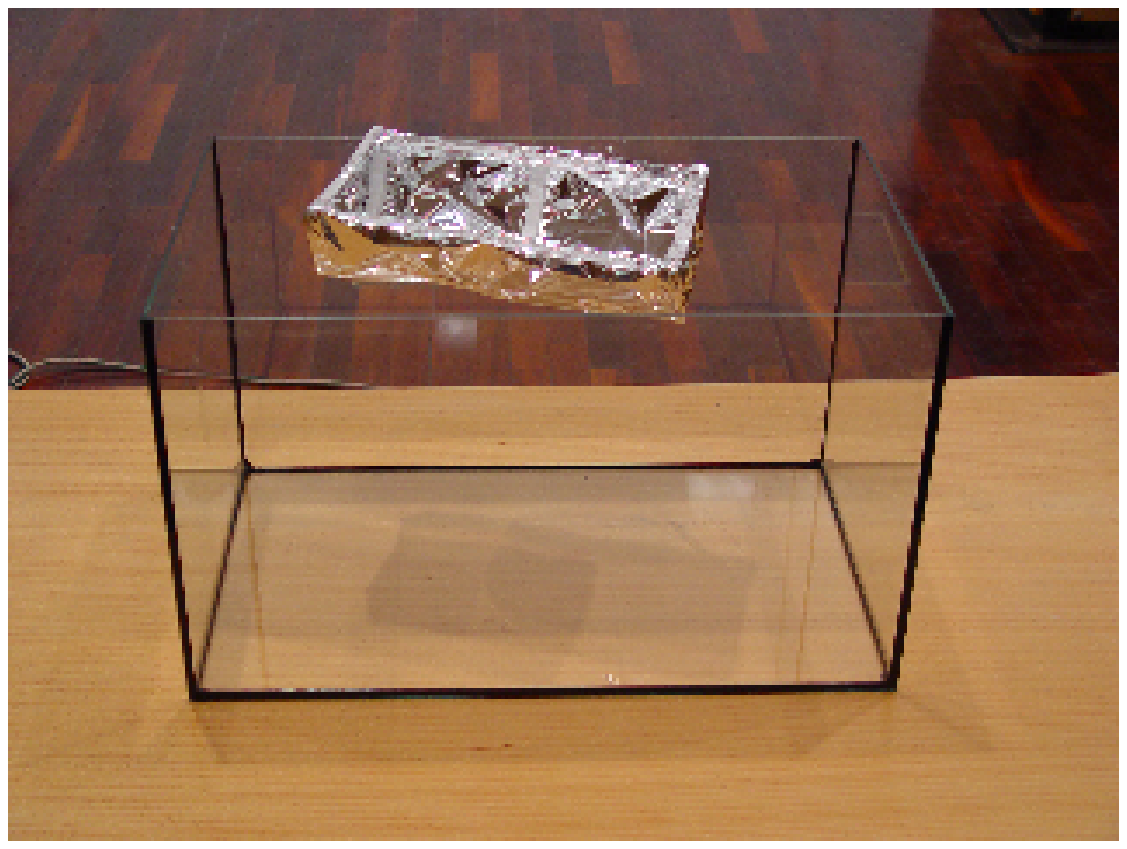}}

\centerline{\small Fig. 1 Aluminium boat floating on sulphur hexafluoride}

\vspace{1.5cm}

\centerline{\includegraphics[width=8.66cm,height=6.5cm,angle=0]{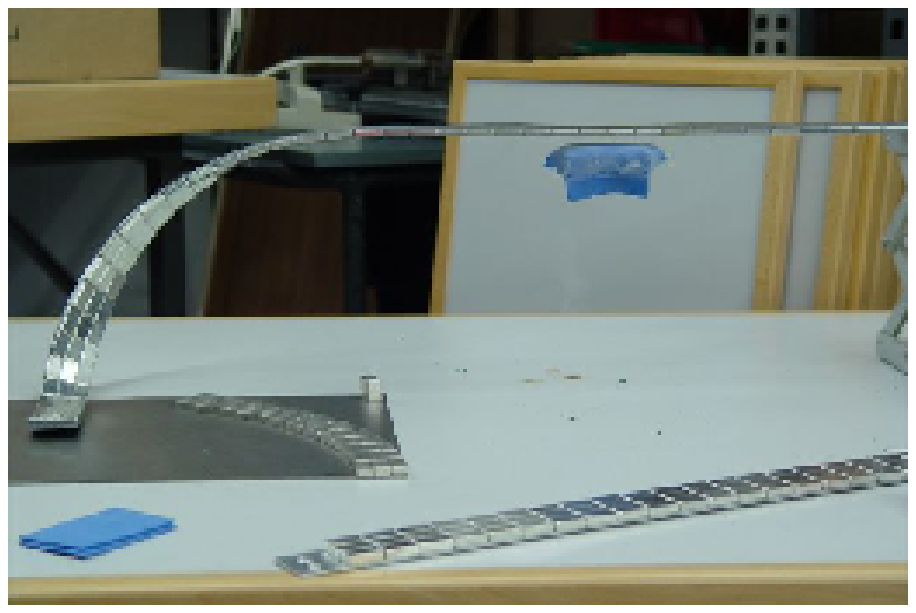}}

\centerline{\small Fig 2. Superconducting train, construction phase, with a blue suspended ``train''.}

\vspace{1.5cm}

\centerline{\includegraphics[width=8.66cm,height=6.5cm,angle=0]{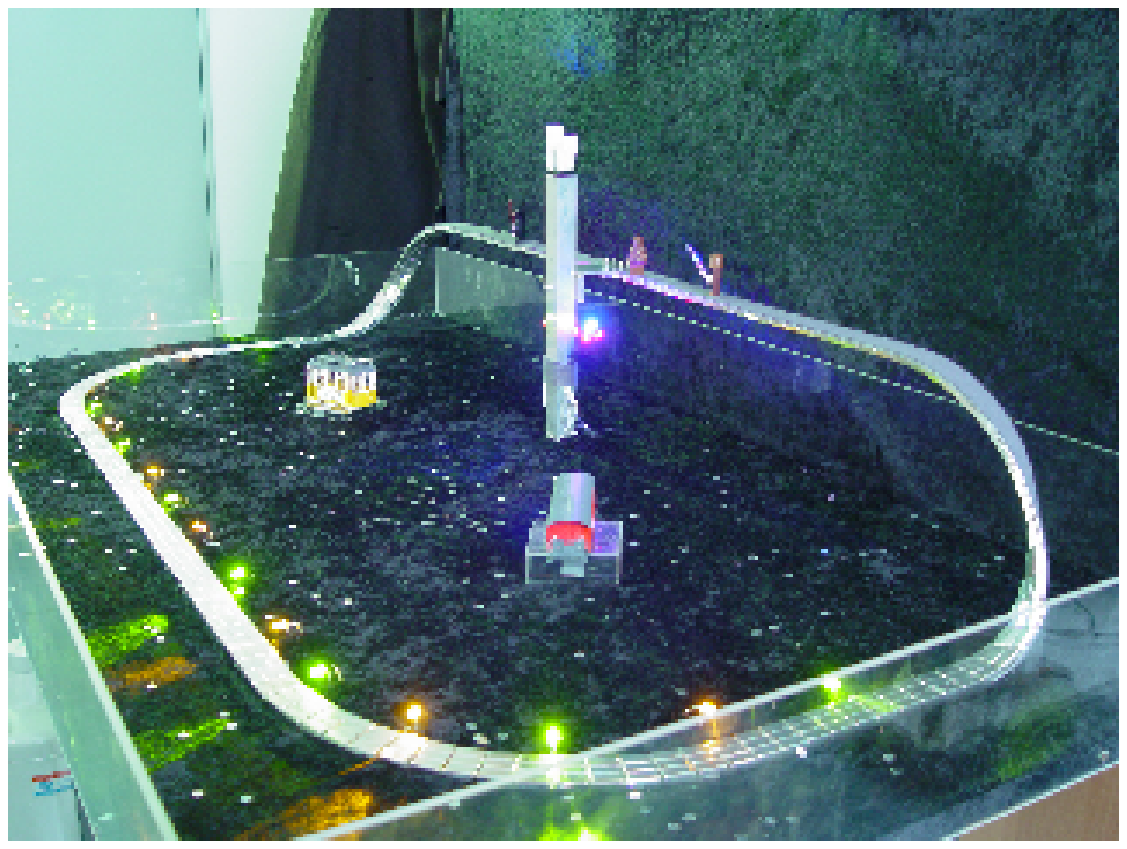}}

\centerline{\small Fig. 3 Superconducting train set with lighting and gondola trains.}


\begin{thebibliography}{99}
\bibitem{german}
{In the German university system physics students
only have classes in physics and mathematics as well as a small minor
requirement such as chemistry or astronomy. The level they have
reached after three years is typically considered equivalent to an
American bachelor degree.  Furthermore, they graduate from high-school
a year later and the men typically have one year of military or
community service. They are therefore also on average one and a half
years older at the same stage.}
\end{thebibliography}
\end{document}